# A holistically 3D-printed flexible millimeter-wave Doppler radar: Towards fully printed high-frequency multilayer flexible hybrid electronics systems

Hong Tang[1], Yingjie Zhang[2], Bowen Zheng[1], Sensong An[1], Mohammad Haerinia[1], Yunxi Dong[1], Yi Huang[1], Wei Guo[2], and Hualiang Zhang[1]

[1]Department of Electrical and Computer Engineering, University of Massachusetts, Lowell, MA, USA

[2]Department of Physics and Applied Physics, University of Massachusetts, Lowell, MA, USA

**Flexible hybrid electronics (FHE) is an emerging technology enabled through the integration of advanced semiconductor devices and 3D printing technology. It unlocks tremendous market potential by realizing low-cost flexible circuits and systems that can be conformally integrated into various applications. However, the operating frequencies of most reported FHE systems are relatively low. It is also worth to note that reported FHE systems have been limited to relatively simple design concept (since complex systems will impose challenges in aspects such as multilayer interconnections, printing materials, and bonding layers). Here, we report a fully 3D-printed flexible four-layer millimeter-wave Doppler radar (i.e., a millimeter-wave FHE system). The sensing performance and flexibility of the 3D-printed radar are characterized and validated by general field tests and bending tests, respectively. Our results demonstrate the feasibility of developing fully 3D-printed high-frequency multilayer FHE, which can be conformally integrated into irregular surfaces (e.g., vehicle bumpers) for applications such as vehicle radars and wearable electronics.**

**Introduction**

Flexible hybrid electronics (FHE), as a new category of electronics, leverages integrated electronics (IC) and emerging 3D printing technologies, therefore featuring a great potential for consumer electronics, industrial electronics, and defense electronics such as wearable electronics[1–4], internet of things (IoT)[5], tele-communications[6–8], and sensing systems[9–11]. By integrating multiple material systems and functional units onto flexible substrates, FHE changes the rigid physical form of traditional electronics in a disruptive innovation, enabling electronic devices to be accommodated to irregular shapes without compromising their functionalities[12]. However, the operating frequencies of most reported FHE are in the lower end of radio frequency (RF) bands[9,13]. Moreover, the implementation of these systems is limited to one or two conductive layers, preventing the development of multilayer wireless systems[14,15] with complex designs.

Meanwhile, millimeter-wave (mmWave) technology is crucial for current/next-generation wireless systems[16,17]. In particular, a mmWave Doppler radar sensor can detect Doppler-based motion, speed, and direction of movement (approaching or leaving) with exceptional speed resolution[18–20]. Its unique features make it suitable for a wide range of applications, including vehicle radars[21], intelligent devices[22], and medicare sensors[23]. For example, it is attractive to integrated mmWave Doppler radar sensors into irregular surfaces (e.g., vehicle door panels and bumpers[24,25]) directly without any other brackets. However, the implementation of commercial mmWave radar sensors generally relies on traditional materials and printed circuit board (PCB) technologies[26], which is rigid in nature with limitations in terms of form-factor.

For next generation wireless systems (e.g., mmWave sensing systems, 5G/6G communication systems), it is envisioned that the combination of FHE and mmWave technology will play a pivotal role. Up to now (despite massive potential of FHE and mmWave technology), there have been few attempts to combine them, especially for complex flexible mmWave systems (e.g., radar sensors). It is primarily due to technical challenges, including interconnections between conductive layers, heat-resistant and flexible printing materials, and reliable bonding layers.

To address these issues, in this work, we demonstrate a fully 3D-printed flexible mmWave Doppler radar by combining FHE and mmWave technology. The proposed radar is composed of four conductive layers, on which mmWave circuits, high-speed digital circuits, power supply circuits and other functional circuits are designed and printed. Stable interconnections among different conductive layers are realized by mechanically drilled blind vias and through vias filled with conductive epoxy. To realize high temperature resistance and good flexibility, a flexible filament is utilized to print insulation layers. The varied geometries of printed insulation layers make it possible to create a radar system with truly 3D structures, enabling more design freedom. Furthermore, to realize robust bonding strength and increase temperature resistances of the printed radar system, high-heat epoxy is coated between each insulation layer to bond them together. To the best of our knowledge, this is the first time an FHE-based mmWave radar (featuring truly 3D structures) has been reported.

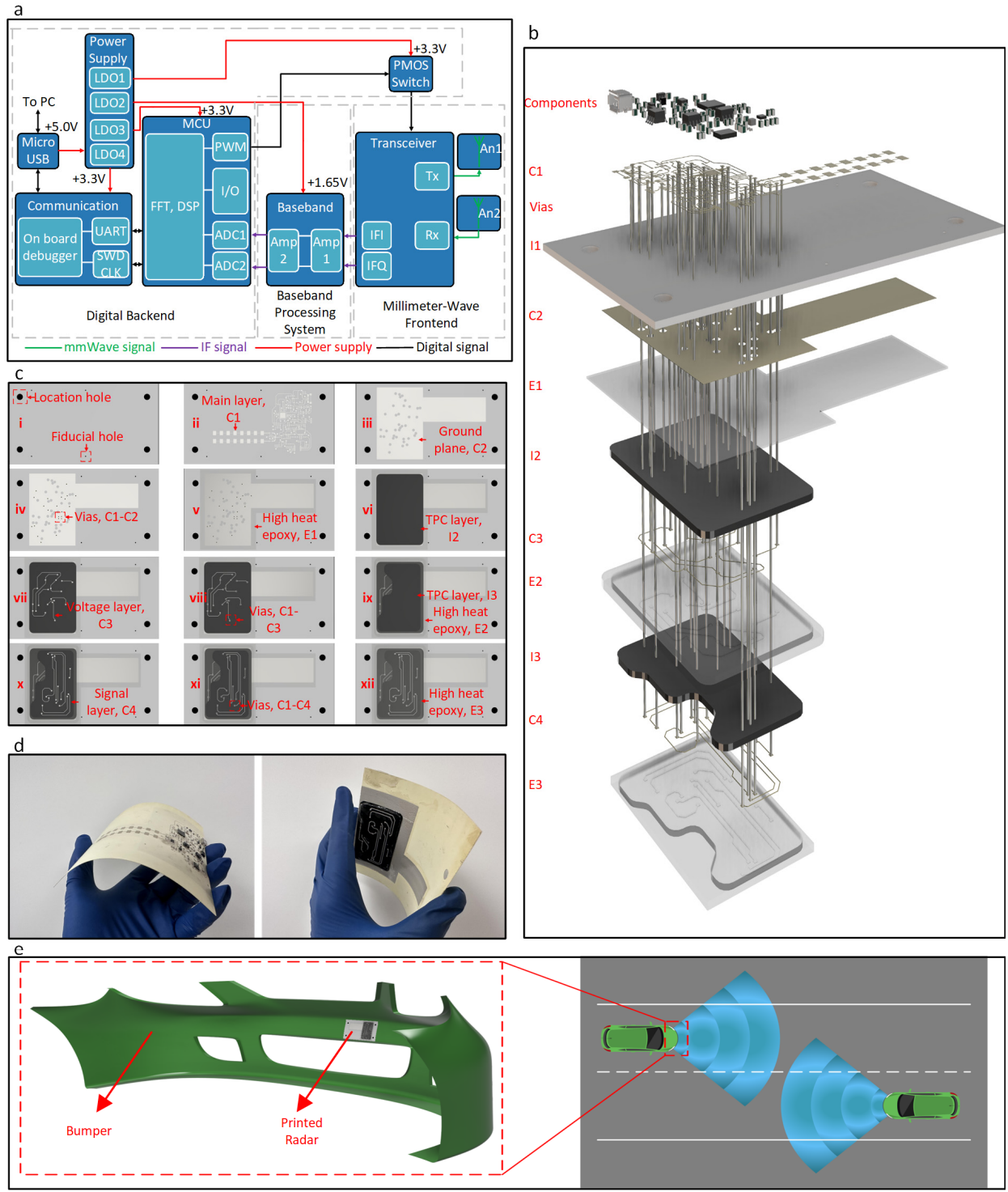

**Fig. 1 | Overview of the proposed system. a,** A system diagram of the proposed radar system. **b,** Layer stack-ups of the proposed fully 3D-printed radar system. **c,** Proposed printing workflows. **d,** Photos of the fabricated radar system under bending. **e,** Schematics of the proposed radar sensor integrated into an automobile bumper.

**Design of the proposed flexible mmWave Doppler radar**

Figure 1a presents the system diagram of the proposed 3D-printed Doppler radar. It is composed of three major units, i.e., the mmWave frontend, the baseband processing system, and the digital backend. The mmWave frontend consists of a highly integrated 24 GHz transceiver IC from Infineon Technologies[27], with peripheral mmWave circuits including a DC block, transmission lines, and two identical 8 × 1 microstrip series-fed patch antenna arrays for transmitting (Tx) and receiving (Rx) channels, respectively. The transceiver generates continuous wave (CW) signals at 24 GHz, and a DC block is applied to filter out DC voltages before entering the antennas (since the CW signals are DC coupled). The filtered CW signals are then radiated by the Tx antenna array. Echoes reflected by a moving target are received by the Rx antenna array and down-converted to intermediate frequency (IF) signals by the transceiver. Depending on the target under test in front of the radar, the IF signals can be very low in amplitude (µV to mV range). To process these low-amplitude signals, the baseband processing system is employed to amplify, filter, and recondition the IF signals. The amplified analog IF signals are transmitted to and digitalized by the digital backend. The digital backend can be further divided into three modules (power supply module, microcontroller unit (MCU) module, and communication module). The power supply module consists of 5 linear and low-dropout regulators (LDOs), which can convert the 5V main supply voltage from an external power supply to required driving voltages for other on-board units and modules with low switching noise. A 12-bit analog-to-digital converter (ADC) integrated on the MCU module samples and converts analog IF signals to digital forms. Additionally, the MCU module is responsible for controlling all other units and modules via serial peripheral interface (SPI) protocol. The communication module supports communication between a PC/laptop and the MCU module so that the raw digital IF data can be captured and further processed on a PC/laptop.

**A fully 3D-printing process for the proposed flexible mmWave Doppler radar**

Figure 1b presents a 3D exploded view of the proposed radar layer stack-up. Specifically, the proposed radar system is composed of four conductive layers (C1, C2, C3, and C4), three insulation layers (I1, I2, and I3), three high-heat epoxy layers (E1, E2, and E3), and vias among conductive layers. All surface-mount devices (SMDs) are assembled on the first conductive layer C1 by conductive epoxy. To realize high-quality conductive interconnections among conductive layers, blind vias (vias between C1-C2 and C1-C3 layers) and through vias (vias between C1-C4 layers) are employed. Since mmWave circuits are printed on the first insulation layer I1, the properties of I1 can significantly influence performances of the developed radar system. To achieve desirable RF performances at mmWave frequency range and ensure flexibility, a 10-mil Rogers RO4350B board is chosen as the first insulation layer[28].

A key challenge in the manufacturing process is the ability to withstand high curing temperatures. This is because the radar board needs to be cured in an oven for several temperature cycles (temperature up to 120°C). Correspondingly, printed insulation layers I2 and I3 are required to withstand such high temperatures. To address this issue, we employed a commercial filament material made of flexible thermoplastic copolyester (TPC). TPC is a highly flexible material with a temperature resistance of up to 190°C, making it an excellent choice for this application (more details about the TPC filament are included in Methods). The geometries of I2 and I3 are customized to implement a multilayer circuit board with 3D structures, as shown in Fig. 1b.

Additionally, the proposed work is challenged by the fact that, although the printed insulation layers can resist high temperatures, the radar board is still susceptible to delamination due to the differences in thermal expansion coefficients (CTE) among the printed layers. To avoid delamination and increase the bonding strength, two high-heat epoxy layers E1 and E2 were applied between each insulation layer to bond the whole radar board. Furthermore, these epoxy layers can prevent printed conductive silver circuits from being exposed to the air and oxidized, therefore extending the durability of the board. Besides, high-heat epoxy was employed to enhance the reliability of vias and improve electrical connections between components and printed circuits during the fabrication (more details are included in Methods).

The developed comprehensive 3D printing process is presented in Fig. 1c. First, four location holes and four fiducial holes were drilled (Fig. 1ci) to align multiple layers. Layer C1 and C2 were printed (Fig. 1cii and Fig. 1ciii) as main circuit layer and ground plane, respectively. Electrical connections between these two layers were achieved by blind vias through Rogers RO4350B with conductive epoxy filling (Fig. 1civ). The first high-heat epoxy layer of 25 um was then coated on the ground plane by a blade coater (Fig. 1cv). The second insulation layer I2 with a total thickness of 300 um (Fig. 1vi) and the third conductive layer C3 (Fig. 1cvii) were printed sequentially. Then, the fabrication processes were repeated for the blind vias for C1-C3 layers (Fig. 1cviii), the second high-heat epoxy layer E2, third insulation layer I3 (Fig. 1cix), fourth conductive layer C4 (Fig. 1cx), and through vias for C1-C4 layers (Fig. 1cxi). Finally, one more high-heat epoxy layer E3 was coated to prevent C4 layer from being exposed to the air and oxidized (Fig. 1cxii) (more details of each layer are presented in Supplementary Figure 1).

**Table 1 | Comparisons with state-of-the-art FHE circuits and systems.**

| Ref. | Fabrication Method | Operating Frequency | Conductive Layer Number | Multilayer Interconnections | 3D Structure |
|---|---|---|---|---|---|
| 8 | Additive Manufacturing | 854 MHz | 2 | No | No |
| 15 | Additive Manufacturing | 24 – 28 GHz | 2 | Yes (contains only one through via) | No |
| 29 | Additive Manufacturing | DC | 1 | No | No |
| 30 | Additive Manufacturing | 13.56 MHz | 2 | No | No |
| 31 | Additive Manufacturing & PCB Technology | DC | 2 | No | No |
| 32 | Additive Manufacturing | DC | 2 | No | No |
| 33 | Additive Manufacturing & PCB technology | DC | 2 | No | No |
| 34 | Additive Manufacturing | 904.4 MHz | 2 | Yes (contains only through vias) | No |
| 35 | Additive Manufacturing | 915 MHz | 1 | No | No |
| This work | Additive Manufacturing & Blade Coating | 24 GHz | 4 (can be further increased) | Yes (contains through vias and blind vias) | Yes |

Once the flexible mmWave board was printed, SMD components were placed and assembled on C1 layer by conductive epoxy. During the printing/assembling process, an electrical inspection was performed after every fabrication step to check for printing failures and fabrication errors (more details of printing processes are presented in Methods).

Figure 1d presents photos of the fully 3D-printed flexible mmWave Doppler radar system under bending. Since the developed Doppler radar system is exceptionally flexible and compact (80 mm * 145 mm), it permits conformal, intimate lamination on curved shapes and surfaces, such as vehicle door panels and bumpers (Fig. 1e).

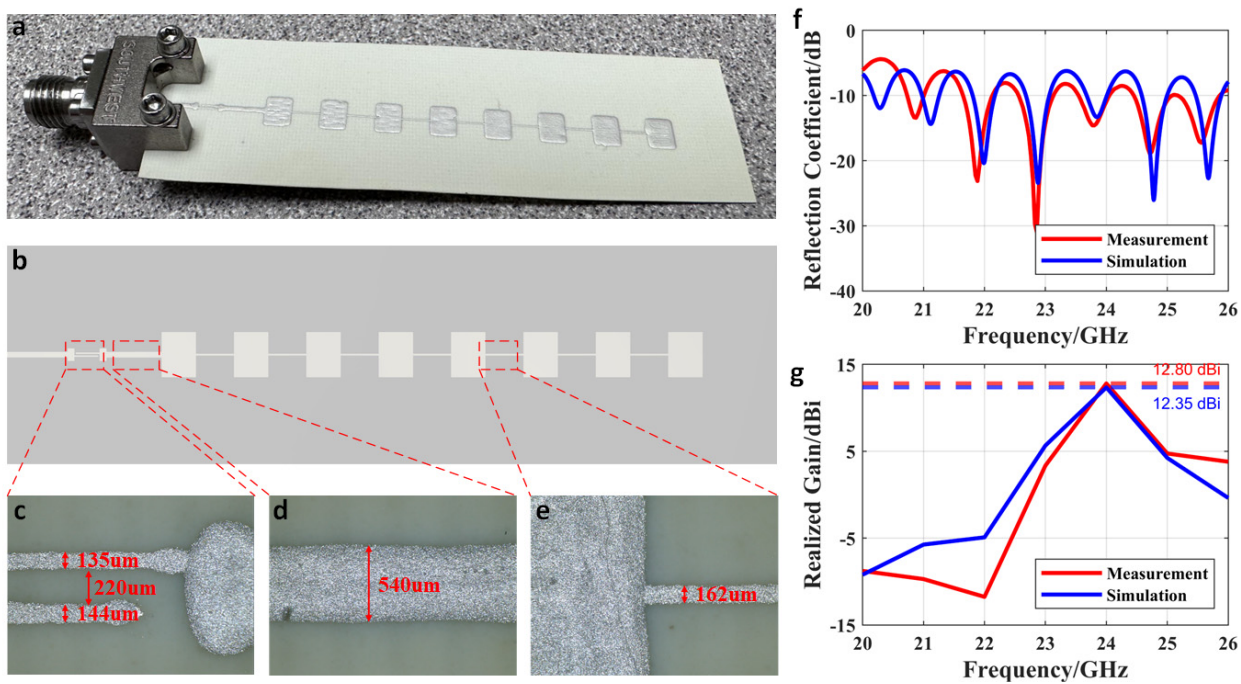

**Fig. 2 | Characterizations of 3D-printed mmWave circuits. a,** A photo of the printed mmWave circuits. **b,** The simulation model of mmWave circuits. **c-e,** Microphotographs of the DC blocker (c), 50-Ohm transmission line (d), and series feeding line (e). **f,** Comparisons between simulated and measured reflection coefficients of mmWave circuits. **g,** Comparisons between the simulated and measured realized gain of mmWave circuits.

Comparisons between our proposed system and state-of-the-art FHE circuits and systems are summarized in Table 1. Compared with contemporary FHE circuits and systems, this work clearly demonstrates a flexible mmWave Doppler radar with multilayer stack-up, stable interconnections (through different vias) and more design freedom (realized by 3D structures). It is worth to mention that our approach can be further developed to increase the number of conductive layers (e.g., 6 or 8 conductive layers) and provide more via options (e.g., blind vias, through vias and buried vias). In general, this approach appears to be promising for the implementation of other high-frequency wireless systems (e.g., communication systems) with FHE.

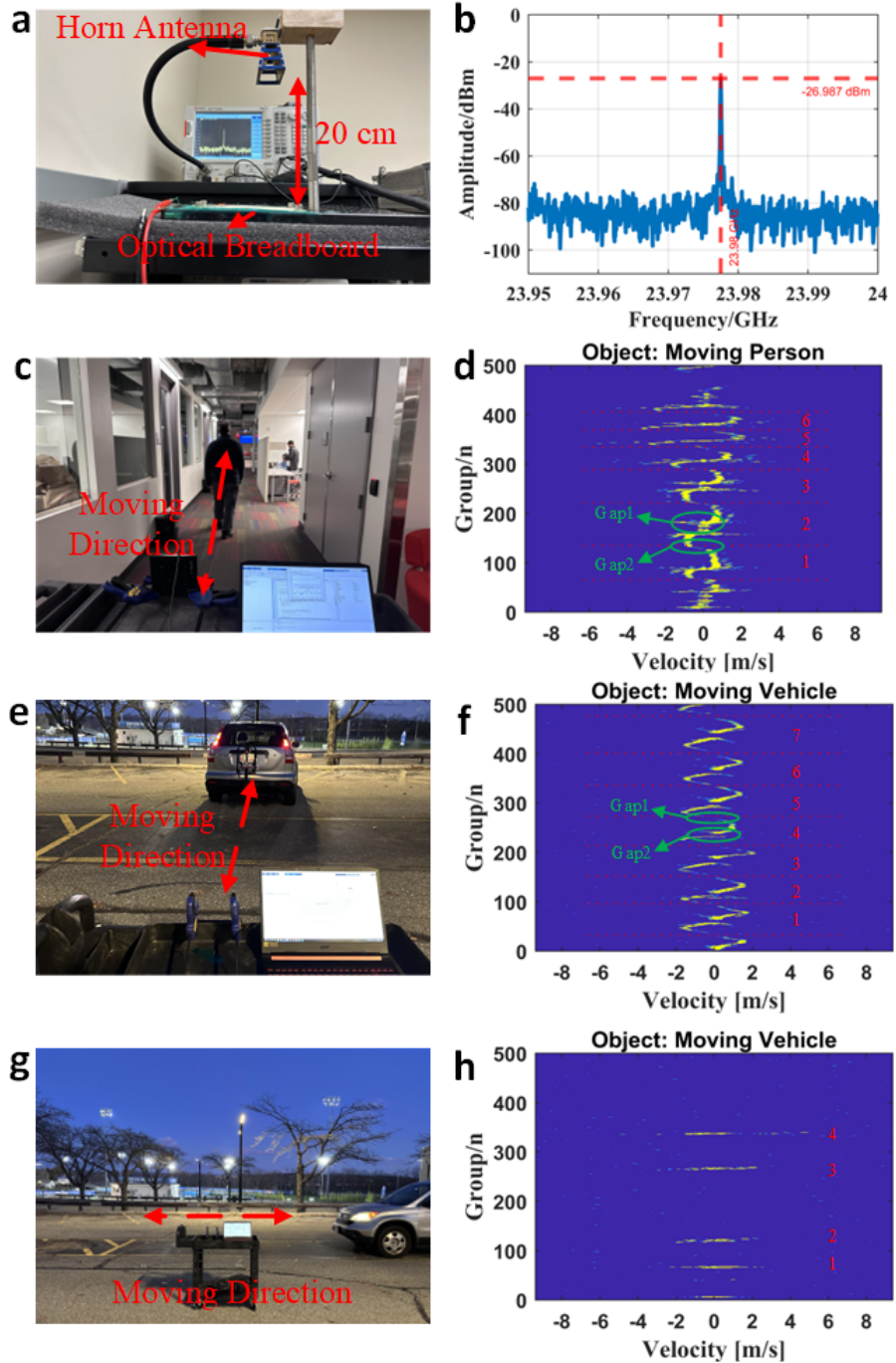

**Fig. 3 | General performance characterization of the radar. a**, Experimental setups for the spectrum test. **b**, Measured spectrum of the printed radar system. **c,** Experimental setups for the indoor field test. **d,** 2D velocity map of field test (c). **e,** Experimental setup (setup I) for the vehicle under test. **f,** 2D velocity map of field test (e). **g,** Experimental setup (setup II) for the vehicle under test. **h,** 2D velocity map of field test (g).

**General performance characterization of the fully printed flexible mmWave Doppler radar**

Performances of the printed mmWave circuits, including the DC block, transmission lines, and 24 GHz microstrip series-fed patch antenna array, are studied in this section (since they are the most crucial parts of the mmWave frontend, and their properties can significantly affect sensing performances of the radar system). A photo and simulation model of the 3D-printed mmWave circuits are presented in Fig. 2a, b. Microphotographs of the printed DC block, 50-ohm transmission line, and patch antenna array series feeding line, as shown in Fig. 2c-e, demonstrate high level of printing accuracy. More details of fabricated and designed parameters of the series-fed patch antenna array are presented in Supplementary Table 1. Figure 2f presents comparisons between simulated and measured reflection coefficients of the printed mmWave circuits. It can be found that good agreements between the simulations and measurements have been realized. Furthermore, simulated and measured antenna array realized gains at broadside from 20 GHz to 26 GHz are presented in Fig. 2g. The realized gain at 24 GHz is 12.80 dBi, which agrees well with the simulation.

Next, a spectrum measurement is carried out to characterize performances of the printed radar system. Figure 3a illustrates the measurement setups. To obtain the output spectrum, the printed radar was fixed on an optical breadboard, and a broadband standard gain horn was placed 20 cm above the radar as a receiver to detect output signals. The measured spectrum is shown in Fig. 3b. It indicates that the actual working frequency of the printed mmWave Doppler radar is around 23.98 GHz (matching with the desired working frequency, i.e., 24 GHz) with a maximum measured power of -26.987 dBm.

To assess detection performances of the 3D-printed Doppler radar system, three field tests were conducted by measuring the moving speed of different targets. During the tests, the printed radar was fixed on a cart by two clamps. A laptop computer was adopted to power the radar and to process the raw IF data received from the radar.

To obtain the moving speed of targets under test, Fourier transform is conducted to determine Doppler frequency shifts of the received IF signals. Relationships between velocity of the moving object and Doppler frequency shifts is given by[18–20,36,37]:

$$f_{Doppler}(Hz) = \frac{2v}{\lambda} \quad (1)$$

where v is velocity of the moving object, $\lambda$ is wavelength of the transmitted mmWave signal, and $f_{Doppler}$ is the Doppler frequency shift. Thus, velocity of the moving object is given by:

$$v = \frac{1}{2} f_{Doppler} \times \lambda \quad (2)$$

According to Equation (2), velocity of the target under test can be determined by the Doppler frequency shift. The processed velocity can be positive or negative. Positive velocity shows that the target under test is departing from the radar, while negative velocity means the target is approaching the radar. An example of a group of received IF raw data and processing progresses are discussed in Methods.

During the field tests, the target under test (i.e., a moving volunteer or a moving vehicle) was required to travel in front of the radar with different moving directions (departing or approaching). When the moving volunteer is under test, a typical indoor test setup[38] is shown in Fig. 3c. During the test, the volunteer was required to move, from 0.5 m to 4 m, radially towards or away from the radar (with an aspect angle of 0°) without carrying any other objects at a normal walking speed to simulate real-life indoor scenarios. The volunteer was instructed to approach the radar first, then departing from the radar to get back to the start point. Figure 3d presents the processed 2D velocity map. X-axis shows the moving speed, while y-axis shows the number of received IF data groups (500 continuous groups of velocity data are presented). As labeled in Fig. 3d, 6 whole moving cycles can be clearly observed. For each cycle, the detected velocity turns negative and then changes to positive, illustrating that the target under test was approaching and departing from the radar. Additionally, we can clearly observe two gaps (i.e., Gap1 and Gap2 as marked in the figure) within each cycle, indicating the target under test stopped and changed moving directions at those moments.

In Fig. 3e-h, two field tests involving a vehicle were conducted. As shown in Fig. 3e, a vehicle under test moved back and forth from 0.5 m to 8 m in front of the radar (similar to the case in Fig. 3c). Figure 3f presents the processed 2D velocity map. As can be seen, the vehicle under test completed 7 cycles within 500 groups of data. Considering that the vehicle under test moved at an almost constant speed, velocity distributions among these cycles are consistent. Since it took longer for the vehicle under test to stop, reverse, and start, gaps (i.e., Gap1 and Gap2 labeled in the figure) in Fig. 3f are more obvious than those in Fig. 3d.

In Fig. 3g, the vehicle under test moved from left to right in front of the radar, and then it reversed its direction and headed back to the left. The minimum distance between the vehicle under test and the radar was 0.5 meters when the vehicle was right in front of the radar. The processed 2D velocity map is presented in Fig. 3h. Four groups of actions (detected velocities changing from negative to positive) can be observed within 500 incessant data groups. More discussions about this field test are presented in Supplementary Figure 3.

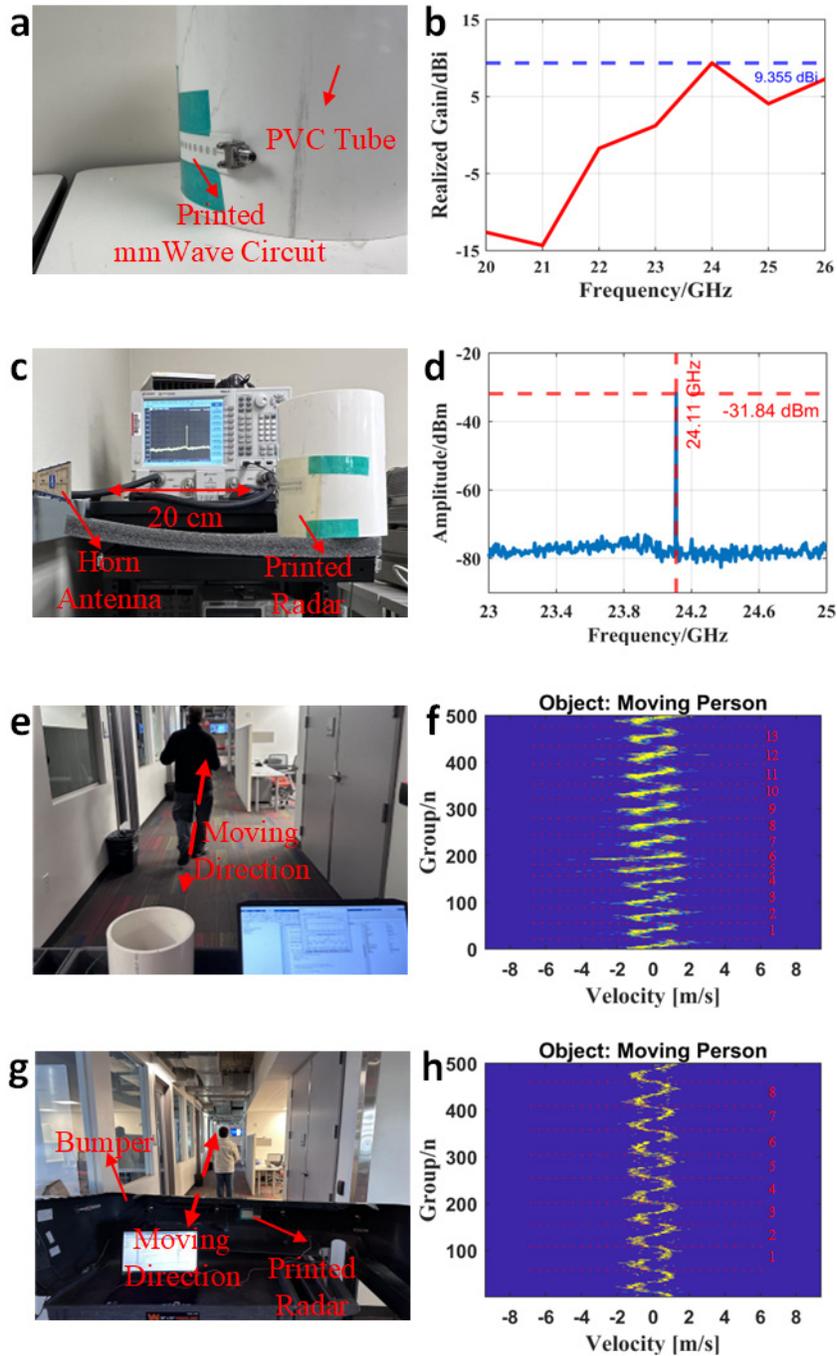

**Fig. 4 | On-object tests of the printed flexible radar. a**, A photo of printed mmWave circuits attached to a PVC tube. **b**, Characterized realized gain of printed mmWave circuits attached on a PVC tube. **c,** Experimental setups for spectrum test. **d,** Measured spectrum of test (d). **e,** Indoor field test setups when the radar was conformally attached to a PVC tube. **f,** 2D velocity map of test (e). **g,** Indoor field test setups when the radar was conformally attached to a vehicle rear bumper. **h,** 2D velocity map of test (g).

Based on the general characterization of the proposed radar, we can conclude that sensing properties of the radar meet our expectations under both indoor and outdoor environments, therefore validating its great potential for applications such as indoor health monitoring systems[39], home security systems[18] and outdoor vehicle speedometers[21].

**On-object tests of the fully printed flexible mmWave Doppler radar**

To demonstrate the flexibility/versatility of the proposed fully 3D-printed radar, it has been integrated with different curved objects representing different application scenarios. In Fig. 4a, a hollow PVC tube was adopted as the platform to test the device under bending[40,41]. The outer diameter of the PVC tube is 168.2 mm, which corresponds to a curvature of 0.012 mm$^{-1}$. As presented, the printed mmWave circuits were conformally attached to the tube outer surface and fixed on the position by tapes. Figure 4b shows the characterized realized gain of bent mmWave circuits at broadside, with a value of 9.355 dBi at 24 GHz. A detailed discussion about the differences between unbent/bent mmWave circuits can be found in Supplementary Figure 4.

To characterize the output spectrum and detection performances of the radar system under bending scenarios, the 3D-printed radar board was conformally attached to the outer surface. The corresponding experimental setups are presented in Fig. 4c, e. In Fig. 4c, a broadband horn antenna was used as a receiver to detect the output signal, and the distance between the radar and the horn antenna was around 20 cm (the experimental setups are similar to Fig. 3a). The measured spectrum of the bent radar is presented in Fig. 4d. It can be noticed that the output frequency of the bent Doppler radar is around 24.11 GHz with a detected magnitude of -31.84 dBm. Compared with the unbent radar, a weakening in the detected power can be observed. We believe these are due to frequency shifts and realized gain decreases of mmWave circuits under bending scenarios, which are further studied in Supplementary Figure 4. In Fig. 4e, an indoor field test was conducted (the experimental setups and motion protocol are similar to Fig. 3c). The measured 2D speed map of a moving volunteer is presented in Fig. 4f. It can be observed that the volunteer approached the radar and departed from the radar for a total of 13 cycles with almost consistent moving velocity, matching the actual test scenarios.

Last but not least, to verify the feasibility of integrating the printed flexible radar into a vehicle, we conformally attached the printed flexible radar to a vehicle rear bumper cover and fixed the radar to the inside of the bumper cover with tapes without use of additional brackets (photos of the radar-on-bumper are presented in Supplementary Figure 5). Then, an indoor field test was conducted. Figure 4g illustrates the experimental setups, and Figure 4h presents the 2D velocity map measured by the radar-on-bumper. Experimentally, we observed that our proposed radar-on-bumper was capable of capturing real-time target velocity information, validating the detection capabilities of our proposed 3D-printed radar in real-world applications.

As a result of the spectrum and field tests, we were able to conclude that the 3D-printed mmWave radar exhibits desirable sensing performance and flexibility and is therefore a very promising technology for developing next-generation wireless systems, such as flexible vehicle radar systems, wearable electronics, and IoT.

Videos of field tests are available in the supplementary.

**Discussions and conclusions**

We have presented a fully 3D-printed flexible Doppler radar, which is realized through the fusion of FHE and mmWave technology. A holistic 3D printing process suitable for multilayer truly 3D

FHE systems has been developed, which enables reliable multilayer interconnections, validates printing materials, and demonstrates robust bonding layers. Based on comprehensive spectrum and field tests, we have demonstrated that the developed fully 3D-printed radar sensor meets our expectations regarding its flexibility and detection abilities. Notably, it can be conformally integrated into vehicle body panels and operate reliably, which has been validated experimentally (e.g., target sensing). Also, the proposed radar can be applied to other wireless sensing and communication systems. Additionally, it is envisioned that the proposed technology combining FHE and mmWave technology can significantly increase the design complexity and operating frequency of future FHE systems. It will pave the way towards developing fully printed multifunctional FHE systems operating at higher frequency bands, such as 77 GHz FMCW radars, mmWave Satcom systems, and wearable wireless systems.

## Methods

**Radar design verifications.** To validate the radar design concept, a four-layer rigid Doppler radar of the same design was fabricated by traditional PCB technologies and tested experimentally. Photos of the fabricated rigid radar board and test results are presented in Supplementary Figure 6 and Supplementary Figure 7.

**Printing materials.** Chemical reagents used in the fabrication process were all commercially available and were used without further purification or modification. All the conductive layers were printed by using commercial silver nanoparticle ink (Novacentrix's Metalon HPS-FG57B ink) during the fabrication. Permatex 84102 (from Permatex) was used as high-heat epoxy. To provide desirable reliability and conductivity for component assembly, a commercial one-part conductive epoxy MG Chemical 9410 was used for assembling the components on printed circuits in low temperatures. This one-part conductive epoxy has comparable electrical conductivity as other two-part epoxies and is more convenient to apply for small soldering pads because of its lower viscosity. For the flexible filament, ID2060-HT high-temperature filament TPC (from DSM Arnitel) was applied.

**3D printing processes and recipes.** The whole 3D printing process is developed based on a nScrypt 3DN-300 printer[42,43] to provide high printing resolution and the capability to print multiple materials. This printer is integrated with three sets of tool heads: Smartpump, nFD, and nMill. The Smartpump tool head was used to print commercial silver nanoparticle inks with high printing resolution. The nFD tool head performs similar functions to a commercial FDM 3D printer, which was used to print flexible high-heat filament TPC. The nMill tool head is a high-speed spindle that was used to drill fiducial holes, position holes, and all the vias.

When drilling location holes, the drill bit diameter was 6 mm. A diameter of 1.0 mm drill bit was used to drill fiducial holes and all vias.

When printing the mmWave frontend, a 75-um pen tip was adopted to realize high printing resolution. To print other circuit traces on C1 layer with faster printing speed, a 100-um pen tip was used. For other conductive layers, i.e., C2, C3, and C3, a 125-um pen tip was employed to print circuit traces.

A 400-um pen tip was used to print insulation layers (i.e., layers I2 and I3). Each insulation layer is composed of two TPC layers. For the first TPC layer, extrusion width and height were configured as 400 um and 200 um, respectively, to improve the first layer adhesion and increase bonding strength. When printing the second TPC layer, the extrusion width and height were configured as 240 um and 100 um. These settings enable a smoother top surface and make the finished top surface finer for printing silver nanoparticle inks. During the printing, bed temperature was set to 50°C, while nozzle temperature was set to 260°C.

During the fabrication, the board was placed in an oven at 120°C for 40 minutes to cure the silver nanoparticle ink after each conductive layer was printed, while the conductive epoxy was cured at 90°C for an hour to realize favorable conductivity. Moreover, high-heat epoxy was cured at room temperature.

**Mechanical strength and reliability enhanced by the high-heat epoxy.** As mentioned, high-heat epoxy was adopted to increase the bonding strength of the printed radar, enhance the reliability of components assembly, and improve the stability of vias.

During the fabrication, nozzle temperature was higher than melting point of the high-heat epoxy. Thus, the extruded TPC material was mixed with melted high-heat epoxy, leading to robust bonding strength between the TPC layer and the epoxy layer. The diagram of TPC extrusion mixed with melted epoxy can be found in Supplementary Figure 8.

In addition to bonding layers, high-heat epoxy was used to enhance the reliability of components assembly[44] on printed circuits. Since the mechanical strength of 1-part conductive epoxy is not so desirable, cracks may occur when the printed radar board is bent, leading to bad electrical connections. To solve this issue, high-heat epoxy was applied to some large chips to attach them on the board tightly and further increase the stability of electrical connections between components and printed circuits. Additionally, for QFP components, their pins were also covered by high-heat epoxy, making the electrical connection hard to fail, even if the printed radar board is bent or under other critical scenarios. The microphotographs of SMD chips with high-heat epoxy are presented in Supplementary Figure 9.

High-heat epoxy was also used to improve the stability of vias. Mechanically drilled vias were filled with conductive epoxy to realize electrical connections. However, the shrinkage effect of conductive epoxy can lead to electrical connection failures[45], especially when the radar board is bent. To resolve this problem, high-heat epoxy was used to fill the cured vias. During the fabrication, since the viscosity of high-heat epoxy before curing is low, it can quickly fill the voids of vias. When high-heat epoxy is cured, solid vias can be realized, and the mechanical strength of vias can be significantly increased. Microphotographs of vias filled with high-heat epoxy are shown in Supplementary Figure 10.

**Simulations.** CST Studio Suite was used to simulate and optimize mmWave circuits (including mmWave antenna arrays, DC blocker, and RF transmission lines). Conductivity of the commercial silver nanoparticle ink is configured as 2M S/m at 24 GHz during the simulation.

**Link budget analysis.** The link budget analysis offers an approximated estimation of the theoretical maximum velocity the radar can detect. In practice, the resulting range will always be smaller. Nevertheless, it is a valuable starting point in characterizing a radar system.

In theory, the Rx antenna received power $P_r$ is given by[38]:

$$P_r = \frac{P_T G_T G_R \lambda^2 \sigma}{(4\pi)^3 R^4} \tag{3}$$

where $P_T$ is transmitted power, $G_T$ and $G_R$ are realized gain for transmitter and receiver antennas, $\lambda$ is wavelength of the transmitted signal, $\sigma$ is radar cross-section (RCS) of the target under test, $R$ is distance between the target and radar. According to the transceiver datasheet, the typical output power $P_T$ is 6 dBm. Considering $\sigma_{human} = 0$ dBsm for a human, $R = 4$ m for indoor test,

Rx received power $P_{r\_human} = -70.4$ dBm. When a vehicle is under test, considering $\sigma_{vehicle} = 6$ dBsm for a vehicle at mmWave range[46] and $R = 8$ m, Rx received power $P_{r\_vehicle} = -76.4$ dBm.

Since the transceiver has a typical noise figure (NF) of 10 dB, and a total receiver bandwidth of $B$ around 1.5 kHz, the receiver thermal noise power floor $P_r$ is given by:

$$P_n = -174\ dBm + NF + 10log_{10}B = -132.2\ dBm \quad (4)$$

When the volunteer is under test, we can estimate the signal-to-noise ratio by:

$$SNR_{human} = P_{r\_human} - P_n = 61.8\ dB \quad (5)$$

When a vehicle is under test, the signal-to-noise ratio is given by:

$$SNR_{vehicle} = P_{r\_vehicle} - P_n = 55.8\ dB \quad (6)$$

Although these SNRs are all calculated based on ideal scenarios excluding all the system losses, they provide enough confidence on radar detection abilities for the volunteer under test within 4 m and vehicle under test within 8 m.

**Algorithms for the Doppler radar.** An example of received IF signals is shown in Supplementary Figure 11a. In-phase parts of the IF signals are shown in red, while quadrature parts are blue. As presented, for each in-phase and quadrature signal group, there are 128 samples in total. To obtain moving speed of the target under test, Fourier transform will be conducted to get the spectrum of received IF signals. The spectrum of received IF signals is presented in Supplementary Figure 11b and a peak at + 99.52 Hz can be observed. According to Equation (2), the processed velocity spectrum based on this IF group is exhibited in Supplementary Figure 11c. We can clearly find a peak when velocity is around 0.61 m/s, illustrating that detected speed of the target under test is 0.61 m/s. According to the sampling theorem that a signal must be sampled at least twice frequency of the original signal, the maximum Doppler frequency shift should be less than half of the sampling rates of ADCs. Thus, the maximum detectable velocity is given by:

$$v_{max} = \frac{1}{4}\ f_{ADC\ sampling\ frequency} \times \lambda \quad (7)$$

Sampling rates of the integrated ADCs are configured as 3000 SPS. Correspondingly, the maximum detectable velocity is 9.38 m/s. It is worth to mention that sampling rates of the ADCs can be modified if faster velocity needs to be detected in the future.

## Data availability

The raw IF data for radar signal processing is provided with this paper.

**Acknowledgments**

We are grateful to Mark W Keene for his fruitful discussion and kind help regarding the radar fabrication and field tests. Additionally, we would like to thank Dr. Chenxi Wang for his assistance with the radar tests.

**Author contributions**

H.T. and H.Z. conceived the idea. H.T. simulated the system. H.T. and Y.Z. conducted the fabrication. H.T. assisted by B.Z., Y.Z., S.A., M.H., Y.D., and Y.H. performed the radar measurement and data analysis. H.Z. supervised the research. All authors contributed to the interpretation of the final results.

**Competing interests**

The authors declare no competing interests.

# A holistically 3D-printed flexible millimeter-wave Doppler radar for self-driving vehicles: Towards fully printed high-frequency multilayer flexible hybrid electronic systems: Supplementary Information

Hong Tang[1], Yingjie Zhang[2], Bowen Zheng[1], Sensong An[1], Mohammad Haerinia[1], Yunxi Dong[1], Yi Huang[1], Wei Guo[2], and Hualiang Zhang[1]

[1]*Department of Electrical and Computer Engineering, University of Massachusetts, Lowell, MA, USA*

[2]*Department of Physics and Applied Physics, University of Massachusetts, Lowell, MA, USA*

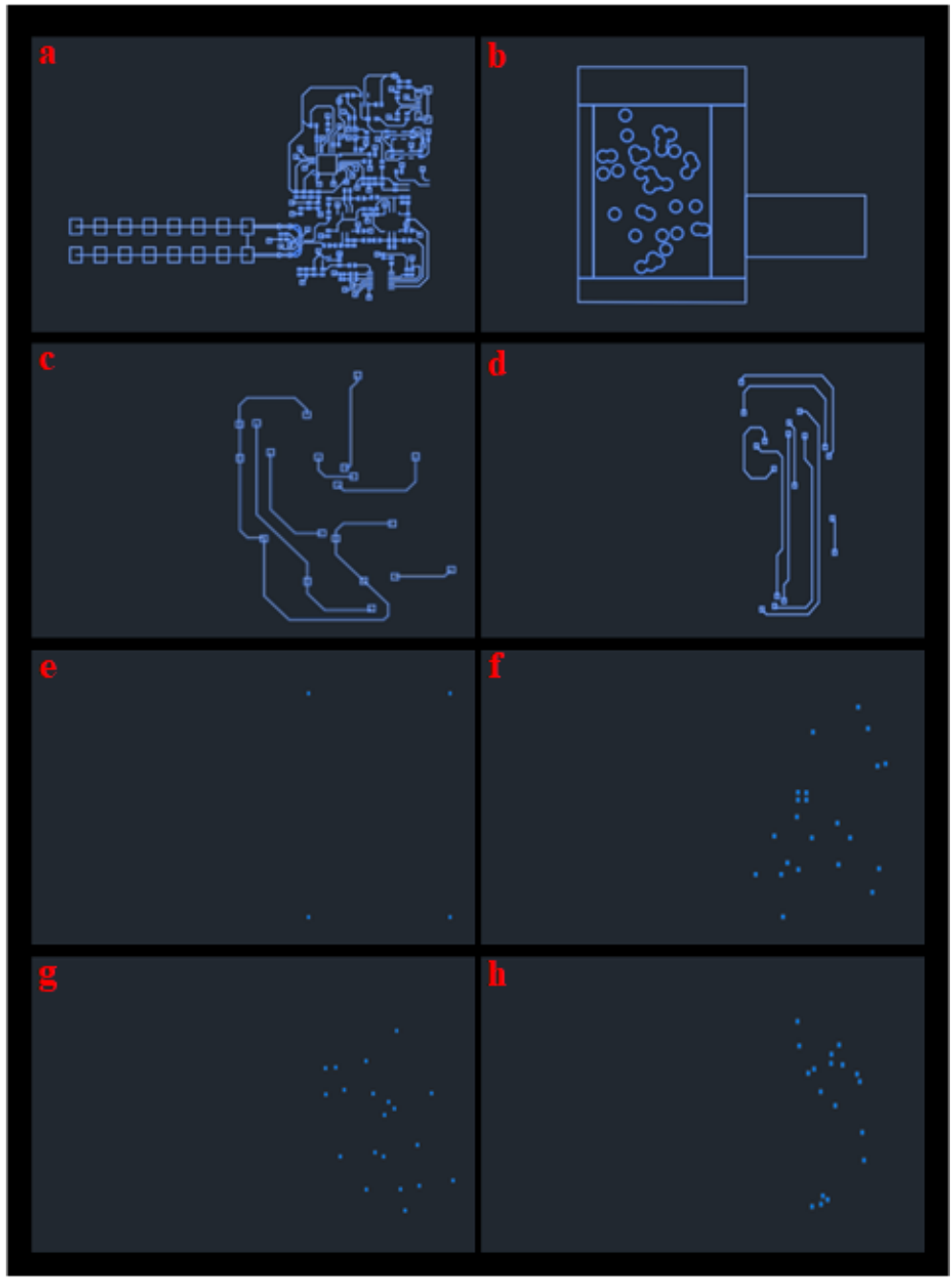

**Supplementary Figure 1. 2D DXF design files for each printed layer. a) C1 layer. b) C2 layer. c) C3 layer. d) C4 layer. e) Fiducial holes. f) C1-C2 vias. g) C1-C3 vias. h) C1-C4 vias.**

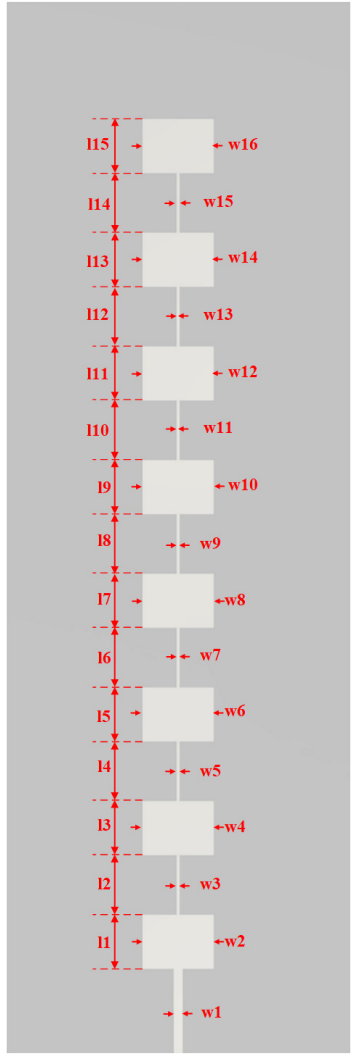

**Supplementary Figure 2. Geometries of the mmWave patch antenna array.**

The geometries of the proposed mmWave patch antenna array are illustrated in Supplementary Figure 2. The design parameters are all labeled in this figure. Moreover, the comparisons between the designed and fabricated parameters are shown in Supplementary Table 1.

**Supplementary Table 1 | Comparisons between the designed and fabricated parameters (labeled in Supplementary Figure 2) of the mmWave patch antenna array.**

| Item Name | Designed Value (mm) | Manufactured Value (mm) | Error Rate | Item Name | Designed Value (mm) | Manufactured Value (mm) | Error Rate |
|---|---|---|---|---|---|---|---|
| w1  | 0.51 | 0.50 | 2.00%  | l1  | 3.12 | 3.10 | 0.64% |
| w2  | 4.1  | 4.13 | 0.73%  | l2  | 3.44 | 3.32 | 3.49% |
| w3  | 0.2  | 0.18 | 10.00% | l3  | 3.12 | 3.14 | 0.64% |
| w4  | 4.1  | 4.11 | 0.24%  | l4  | 3.44 | 3.26 | 5.23% |
| w5  | 0.2  | 0.19 | 5.00%  | l5  | 3.12 | 3.11 | 0.32% |
| w6  | 4.1  | 4.08 | 0.49%  | l6  | 3.44 | 3.26 | 5.23% |
| w7  | 0.2  | 0.19 | 5.00%  | l7  | 3.12 | 3.14 | 0.64% |
| w8  | 4.1  | 4.08 | 0.49%  | l8  | 3.44 | 3.29 | 4.36% |
| w9  | 0.2  | 0.21 | 5.00%  | l9  | 3.12 | 3.11 | 0.32% |
| w10 | 4.1  | 4.11 | 0.24%  | l10 | 3.44 | 3.32 | 3.49% |
| w11 | 0.2  | 0.18 | 10.00% | l11 | 3.12 | 3.12 | 0.00% |
| w12 | 4.1  | 4.09 | 0.24%  | l12 | 3.44 | 3.29 | 4.36% |
| w13 | 0.2  | 0.21 | 5.00%  | l13 | 3.12 | 3.09 | 0.96% |
| w14 | 4.1  | 4.13 | 0.73%  | l14 | 3.44 | 3.31 | 3.78% |
| w15 | 0.2  | 0.18 | 10.00% | l15 | 3.12 | 3.11 | 0.32% |
| w16 | 4.1  | 4.10 | 0.00%  |     |      |      |       |

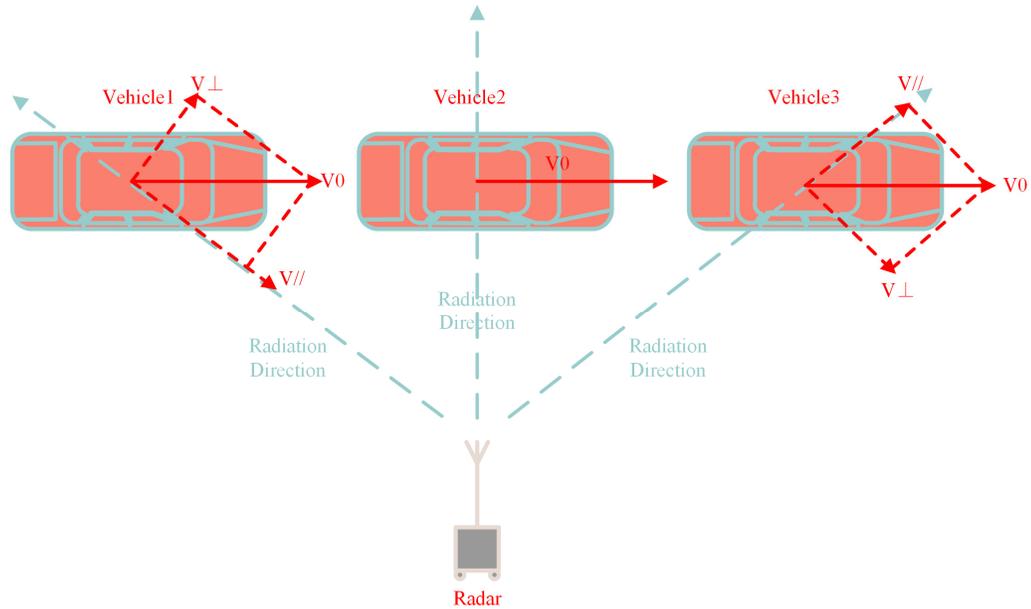

**Supplementary Figure 3. Velocity diagram for the moving vehicle under test.**

The velocity diagram for the moving vehicle is shown in Supplementary Figure 3. For vehicle1, it is approaching the radar main beam. Since the moving direction is not in parallel with the radar radiation direction, the velocity needs to be decomposed into two components, the parallel component V// and the perpendicular component V⊥. However, only the parallel component V// can be detected by the radar. It can be observed that the detectable velocity component V// is opposite to the radar radiation direction. Thus, the detected velocity is negative. When the vehicle is right in front of the radar main beam (i.e., vehicle2 shown in the figure), the detected velocity is zero since there is no parallel velocity component V//. When the vehicle departs from the radar (i.e., vehicle3 shown in the figure), the detectable velocity component V// is in the same direction as the radar radiation direction. Thus, the detected velocity is positive in this case.

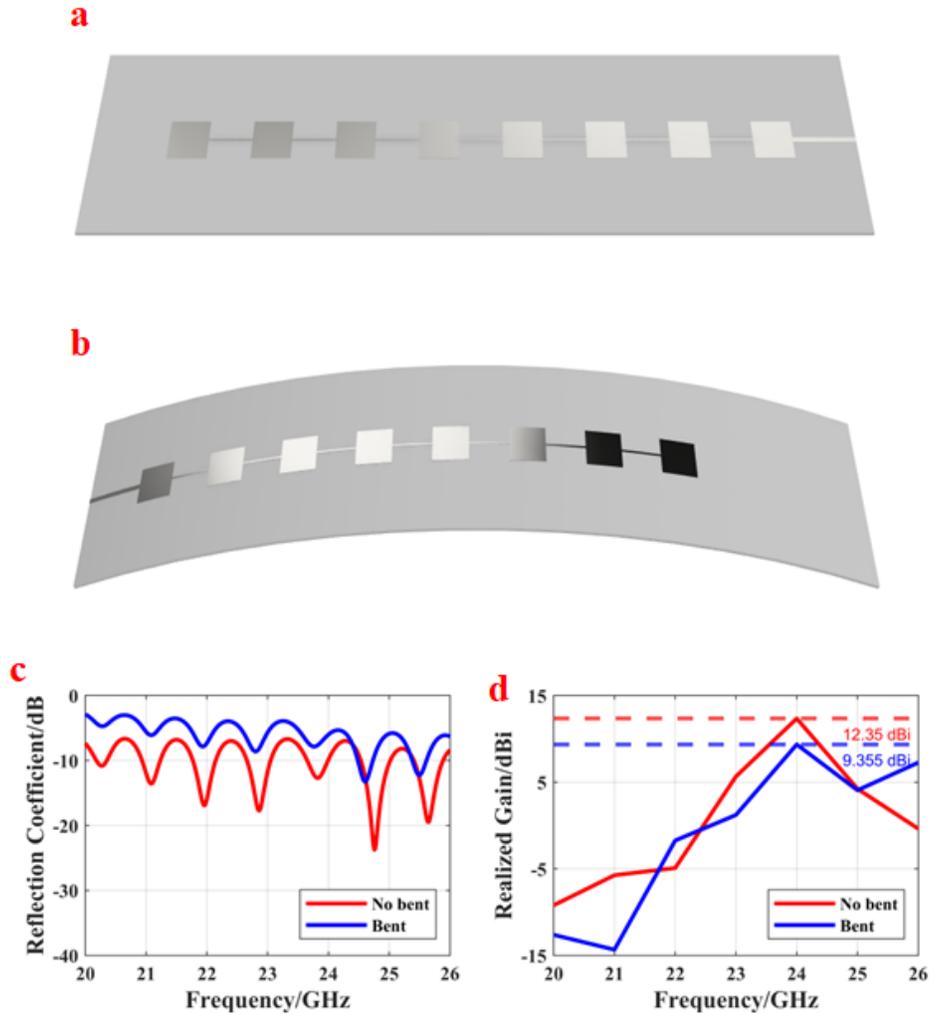

**Supplementary Figure 4. a) Simulation model of the un-bent antenna array. b) Simulation model of the bent antenna array. c, d) Simulated reflection coefficients and realized gains of the bent and un-bent antenna arrays.**

In Supplementary Figure 4b, the antenna array is bent with a radius of 84.1 mm (the size is identical to the radius of the PVB tube during the bending tests) and simulated in CST with a discrete port. Simulated responses of the bent and un-bent antenna arrays are presented in Supplementary Figure 4c and Supplementary Figure 4d, respectively. It can be observed that the radiation performance of the antenna array will change when it is bent. The reflection coefficient will become worse when the antenna is bent, resulting in poor impedance matching at 24 GHz. Correspondingly, the realized gain will decrease by 3 dBi when the array is bent. As a result, decrease in realized gain results in a decrease in the magnitude of the output spectrum (i.e., presented in Fig. 4d in the main manuscript).

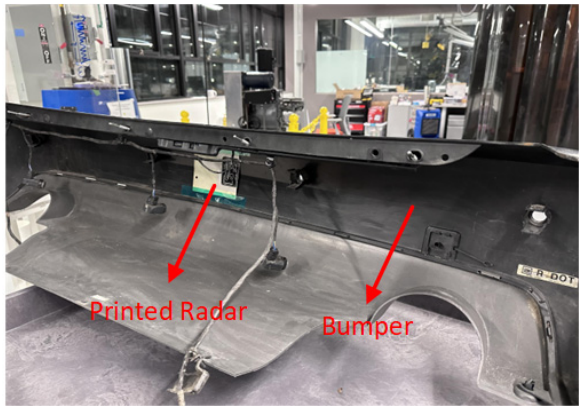 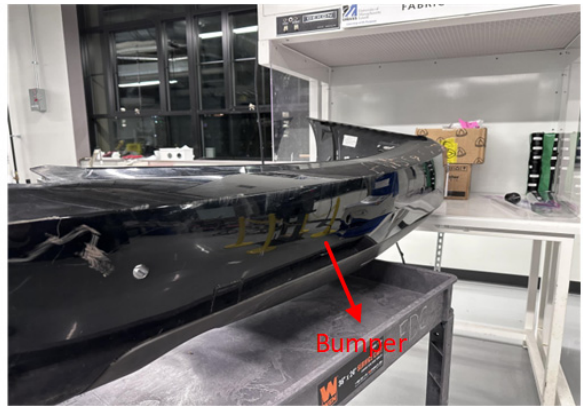

**Supplementary Figure 5.** Photos of the printed flexible radar which is conformally integrated with a vehicle rear bumper.

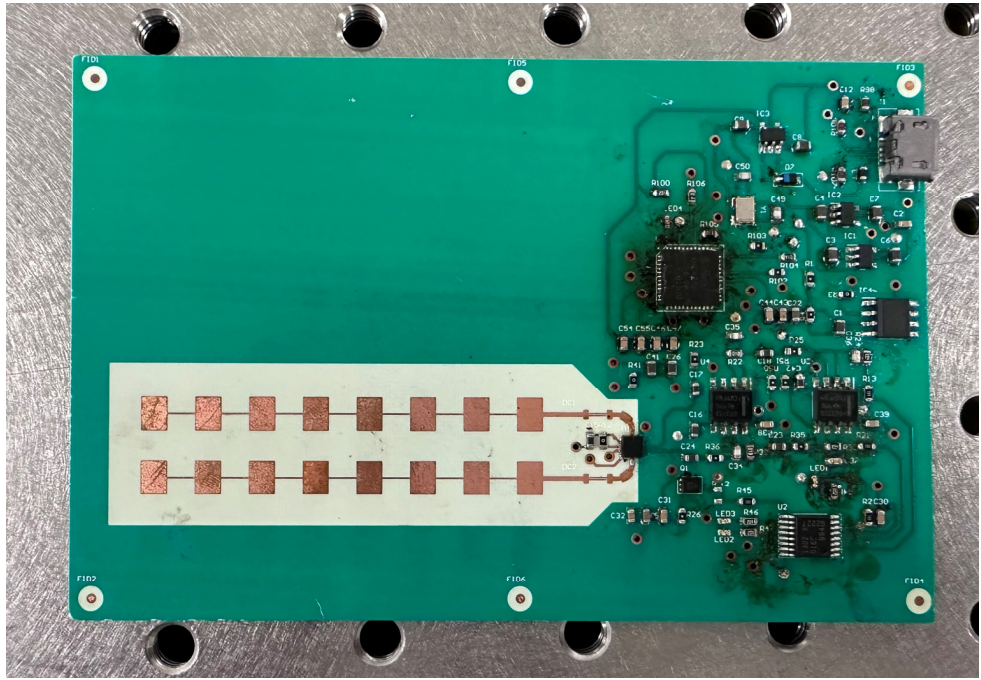

**Supplementary Figure 6. A photo of the rigid Doppler radar manufactured by conventional PCB technologies.**

The photo of a rigid Doppler radar is shown in Supplementary Figure 6. It follows almost the same design as our proposed 3D printed Doppler radar, except for the customized shapes of the insulation layers I2 and I3 shown in Fig. 1b of the main manuscript. Additionally, all the vias are designed as through vias to save the cost for this rigid board. This rigid board is manufactured to validate the design concept of our radar board. The performance of this rigid Doppler radar is presented in Supplementary Figure 7.

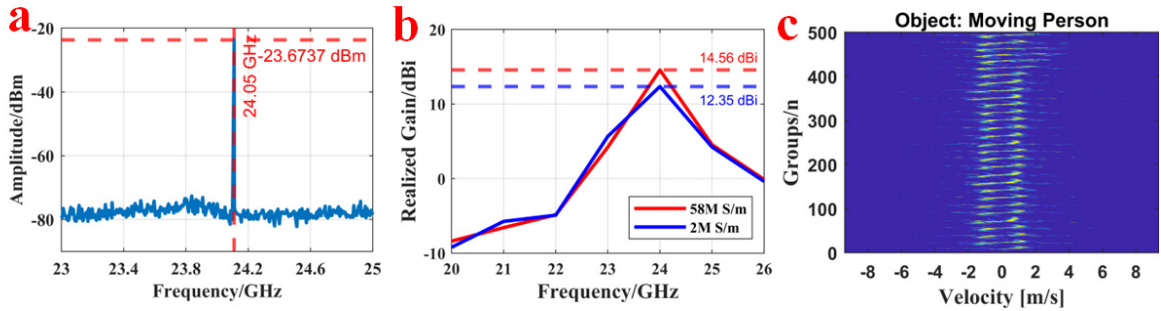

**Supplementary Figure 7. Performance of the rigid Doppler radar. a) Measured output spectrum of the rigid radar. b) Simulated realized gain with different conductivities. c) Field tests results of the rigid radar.**

The experimental setups are identical to the tests for 3D-printed radar. The output spectrum of the RF module is shown in Supplementary Figure 7a. It can be seen that the output frequency of the rigid radar is 24.05 GHz, and the detected output power at that frequency is -23.6737 dBm. It is worth to mention that the measured output amplitude is 3.31dBm larger than that of the printed radar. We believe that it is mainly because the conductivity of the commercial silver ink, which is around 2M S/m, is smaller than that of the copper, which is 58M S/m. Compared with copper, the lower conductivity of commercial silver ink leads to lower efficiency, lower realized gain, and lower output spectrum magnitude. Simulated realized gains with different values of conductivities are presented in Supplementary Figure 7b. It can be observed that a lower conductivity will lead to a lower gain. Indoor field tests were conducted to characterize performance of the rigid radar. The processed 2D velocity mapping is shown in Supplementary Figure 7c.

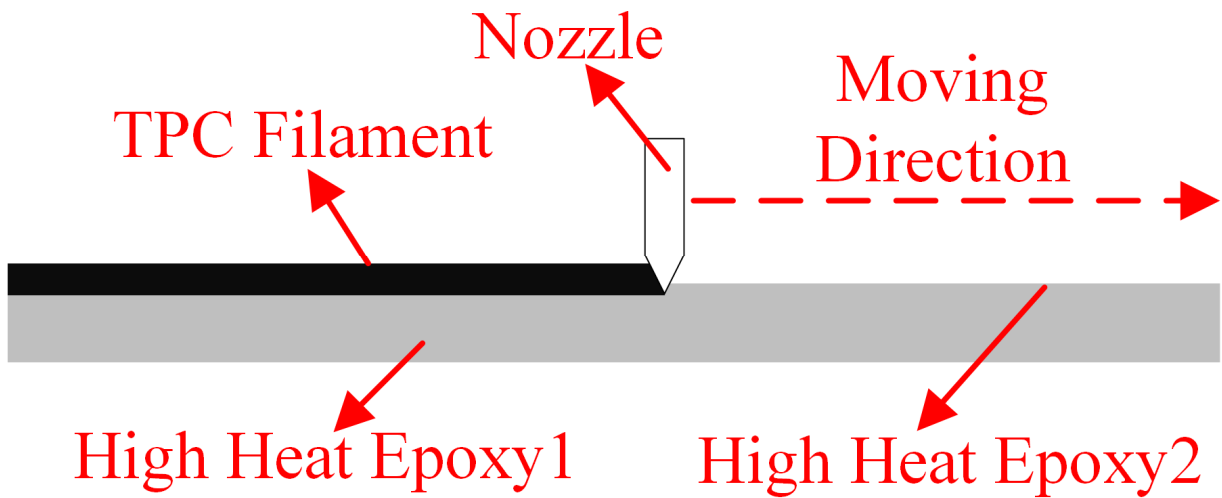

**Supplementary Figure 8.** Diagram for melt high-heat epoxy mixed with printed TPC

The Diagram for the melted high-heat epoxy mixed with the printed TPC layer is shown in Supplementary Figure. 8. Since temperature of the printing nozzle is much higher than melting point of the high-heat epoxy, the top part of the epoxy layer (e.g., high heat epoxy2) will be melted and mixed with the extruded TPC layer to increase bonding strength between the TPC layer and the epoxy layer. A video is available in supplementary videos to further illustrate the mixing processes.

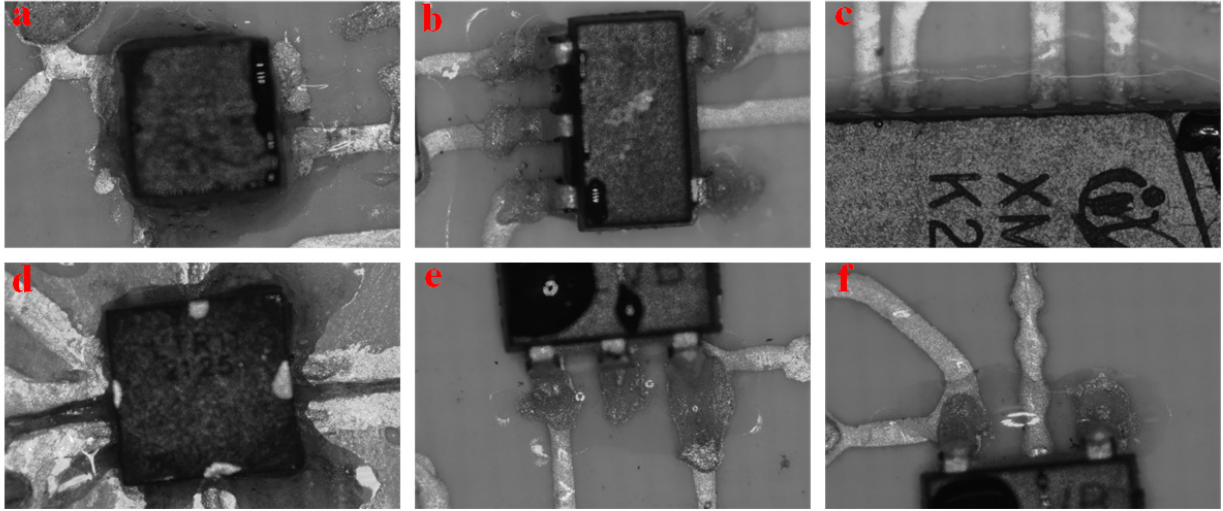

**Supplementary Figure 9. The microphotographs of the SMD chips/components with high-heat epoxy to increase the reliability of the radar system. a) CMOS switch. b) LDO. c) MCU. d) mmWave transceiver. e) LDO. f) LDO.**

The major SMD chips, such as CMOS switch (shown in Supplementary Figure 9a), LDOs (shown in Supplementary Figure 9b, Supplementary Figure 9e, and Supplementary Figure 9f), MCU (shown in Supplementary Figure 9c), 24 GHz transceiver (shown in Supplementary Figure 9d), are all coated with transparent high-heat epoxy to further enhance the reliability. Without high-heat epoxy, the electrical connection may fail since the mechanical strength of the cured conductive epoxy is not strong enough to maintain good connections, especially when the printed radar is bent. When the high-heat epoxy is applied, the chips will be attached to the board tightly and maintain a good electrical connection, even if the board is bent or under other scenarios.

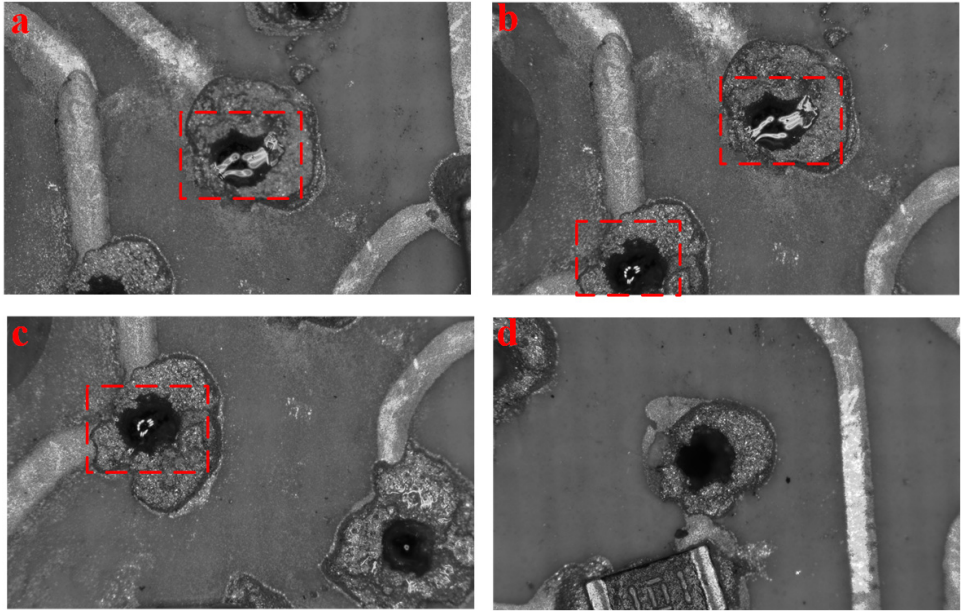

**Supplementary Figure 10. Microphotograph of the vias. a-c) Vias filled with high-heat epoxy. d) Vias without high-heat epoxy.**

The vias filled with high-heat epoxy are displayed in Supplementary Figure 10a-c, while a via without high-heat epoxy is shown in Supplement Figure 10d. In Supplementary Figure 10a-c, the vias filled with conductive epoxy are cured at 90°C for 60 minutes, then filled with high-heat epoxy. Since the viscosity of the epoxy is low before curing, the high-heat epoxy can fill the inner holes of the vias quickly, resulting in a more robust via connection when the high-heat epoxy is fully cured. The cured high-heat epoxy is highlighted in the figure.

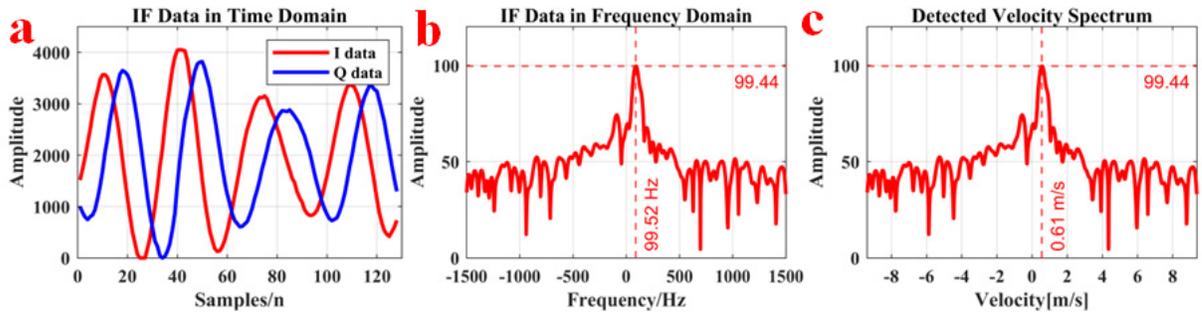

**Supplementary Figure 11. An example of the raw IF data before and after processing. a) The received raw IF data. b) The spectrum of the IF data. c) The processed velocity spectrum.**